\documentclass[11pt]{article}

\usepackage[final]{acl}
\usepackage{times}
\usepackage{latexsym}
\usepackage{makecell}
\usepackage{amsmath}
\usepackage[T1]{fontenc}
\usepackage[utf8]{inputenc}
\usepackage{microtype}
\usepackage{inconsolata}
\usepackage{graphicx}
\usepackage{subfigure}
\usepackage{enumitem}
\usepackage{booktabs}
\usepackage{hyperref}
\usepackage{caption}
\usepackage{marvosym}
\usepackage{colortbl}
\usepackage{xcolor}
\usepackage{multirow}
\usepackage{siunitx}
\usepackage{listings}
\usepackage{setspace}

\definecolor{mygreen}{RGB}{0, 128, 0}
\definecolor{highlight}{rgb}{0.93, 0.95, 1.0} 
\definecolor{zebra}{gray}{0.95}

\definecolor{graybg}{rgb}{0.95, 0.95, 0.95}

\definecolor{red1}{rgb}{1.0, 0.9, 0.9}
\definecolor{red2}{rgb}{1.0, 0.7, 0.7}
\definecolor{red3}{rgb}{1.0, 0.5, 0.5}
\definecolor{red4}{rgb}{1.0, 0.3, 0.3}
\definecolor{codeblue}{RGB}{40, 100, 200}
\definecolor{codegray}{RGB}{120, 130, 140}
\definecolor{codeline}{RGB}{230, 230, 230}

\definecolor{vscodelight-keyword}{RGB}{0, 0, 255}      % 蓝色: if, for, def, class
\definecolor{vscodelight-string}{RGB}{163, 21, 21}    % 深红: "strings"
\definecolor{vscodelight-comment}{RGB}{0, 128, 0}     % 绿色: # comments
\definecolor{vscodelight-function}{RGB}{121, 94, 38}  % 棕色: 方法调用
\definecolor{vscodelight-class}{RGB}{38, 127, 153}    % 青蓝色: Manim 类名
\definecolor{vscodelight-number}{RGB}{9, 134, 88}     % 绿松石色: 数字
\definecolor{vscodelight-gray}{RGB}{128, 128, 128}    % 灰色: 行号
\definecolor{vscodelight-bg}{RGB}{255, 255, 255}      % 背景: 纯白
\definecolor{diffgreen}{RGB}{0, 100, 0}
\definecolor{diffred}{RGB}{160, 0, 0}

\newcommand\footnoteONLYtext[1]{
    \let \mybackup \thefootnote
    \let \thefootnote \relax
    \footnotetext{#1}
    \let \thefootnote \mybackup
    \let \mybackup \imareallyundefinedcommand}

\title{\hspace{-0.4em}
\raisebox{-0.3\height}{\includegraphics[height=1.4cm]{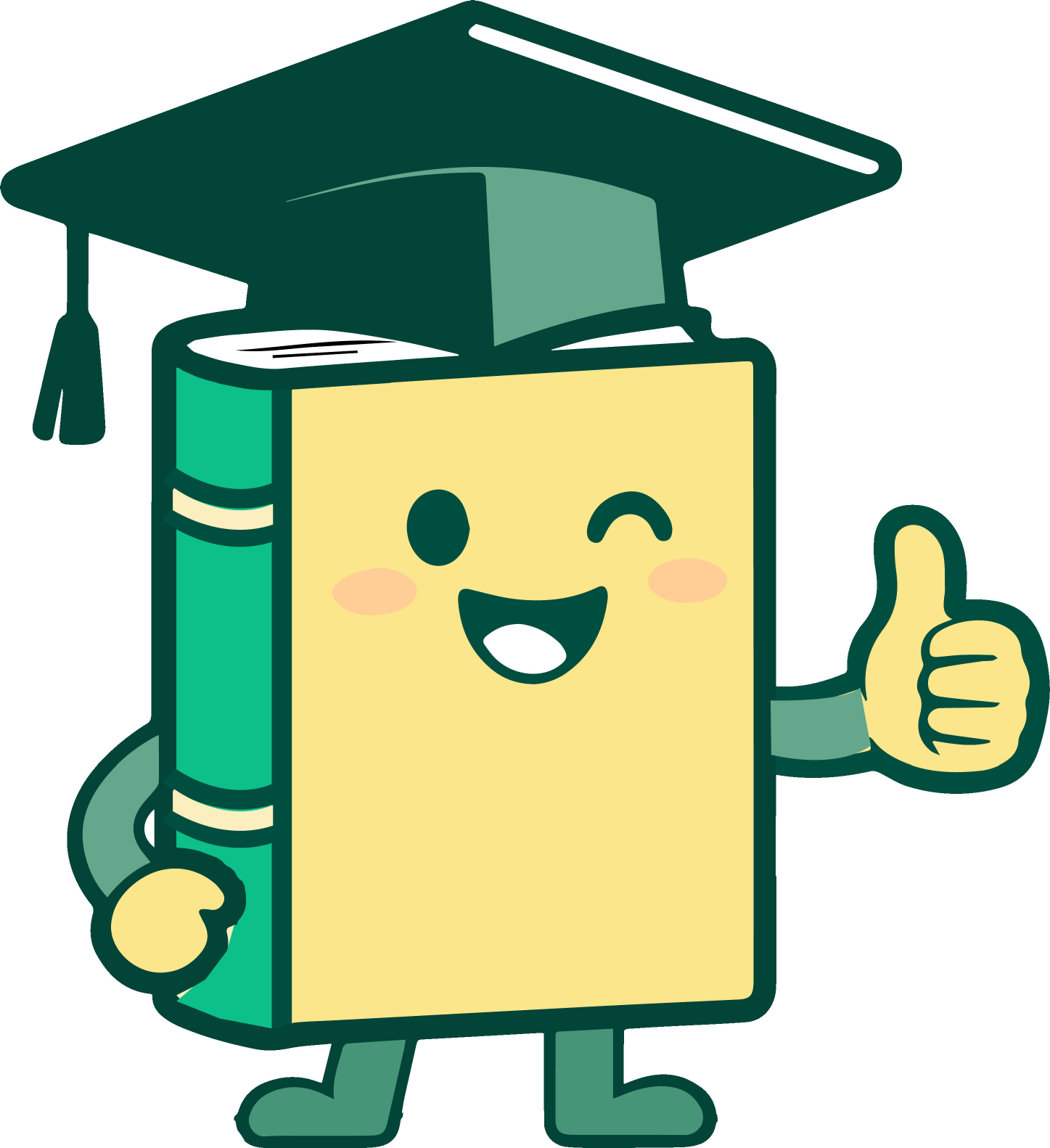}} \hspace{0.1em}
\textbf{TeachMaster: Generative Teaching via Code}
}

\author{
Yuheng Wang, Runde Yang, Lin Wu, Jie Zhang, Jingru Fan, Tianle Zhou,\\
\textbf{Ruoyu Fu, Huatao Li, Ruijie Shi, Siheng Chen, Weinan E, Chen Qian\textsuperscript{\Letter}}\\
\includegraphics[height=0.30cm]{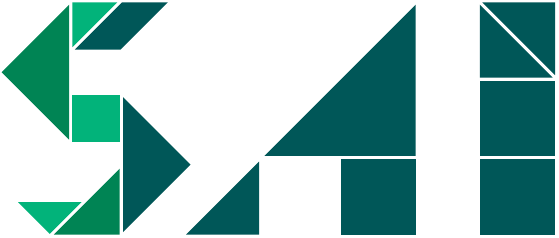}
School of Artificial Intelligence, Shanghai Jiao Tong University\\
\texttt{yuhengwangagent@gmail.com \quad qianc@sjtu.edu.cn}\\
\url{www.teachmaster.cn}
}

\begin{document}

\twocolumn[{
    \renewcommand\twocolumn[1][]{#1}
    \maketitle
}]

\begin{abstract}
The scalability of high-quality online education is hindered by the high costs and slow cycles of manual content creation.
Despite advancements in video generation, current approaches often fail to ensure pedagogical structure and precise control due to their pixel-level, black-box nature.
In this paper, we propose Generative Teaching, a novel paradigm shifting educators from manual creators to high-level directors who focus on pedagogical intents while agents handle the execution. To realize this vision, we introduce TeachMaster, a multi-agent framework that leverages code as an intermediate semantic medium. Unlike traditional video generation methods, TeachMaster orchestrates a collaborative team of agents, spanning planning, design, and rendering, to automate the production of interpretable, editable, and curriculum-ready educational videos. 
Experiments validate that TeachMaster significantly boosts production efficiency without compromising structural coherence or visual fidelity, slashing production costs to only 0.3\% of traditional online course videos and providing a robust solution for scalable education.
\end{abstract}

\section{Introduction}

The global education system faces significant challenges, including the uneven distribution of high-quality educators~\cite{stromquist2018global}, insufficient personalization~\cite{23235ee3-4daf-3b9d-917d-e7a00b919bf0,kasneci2023chatgpt}, and lagging content updates~\cite{antoninis2023global, meng2024scientometric}, all of which limit equitable access to learning opportunities. Although the internet has enabled the widespread dissemination of digitized courses, mainstream online education platforms such as Coursera and edX remain largely confined to the distribution of pre-recorded material~\cite{Reich2019TheMP}. Content creation heavily relies on manual design, production, modification, and recording~\cite{xalxo2025online,Guo2014HowVP}, a process characterized by high production costs (averaging approximately \$30,000 per course), slow update cycles, and limited scalability~\cite{hollands2014resource,raccoongang2025cost}.

\begin{figure}[t]
    \centering
    \includegraphics[width=0.99\linewidth]{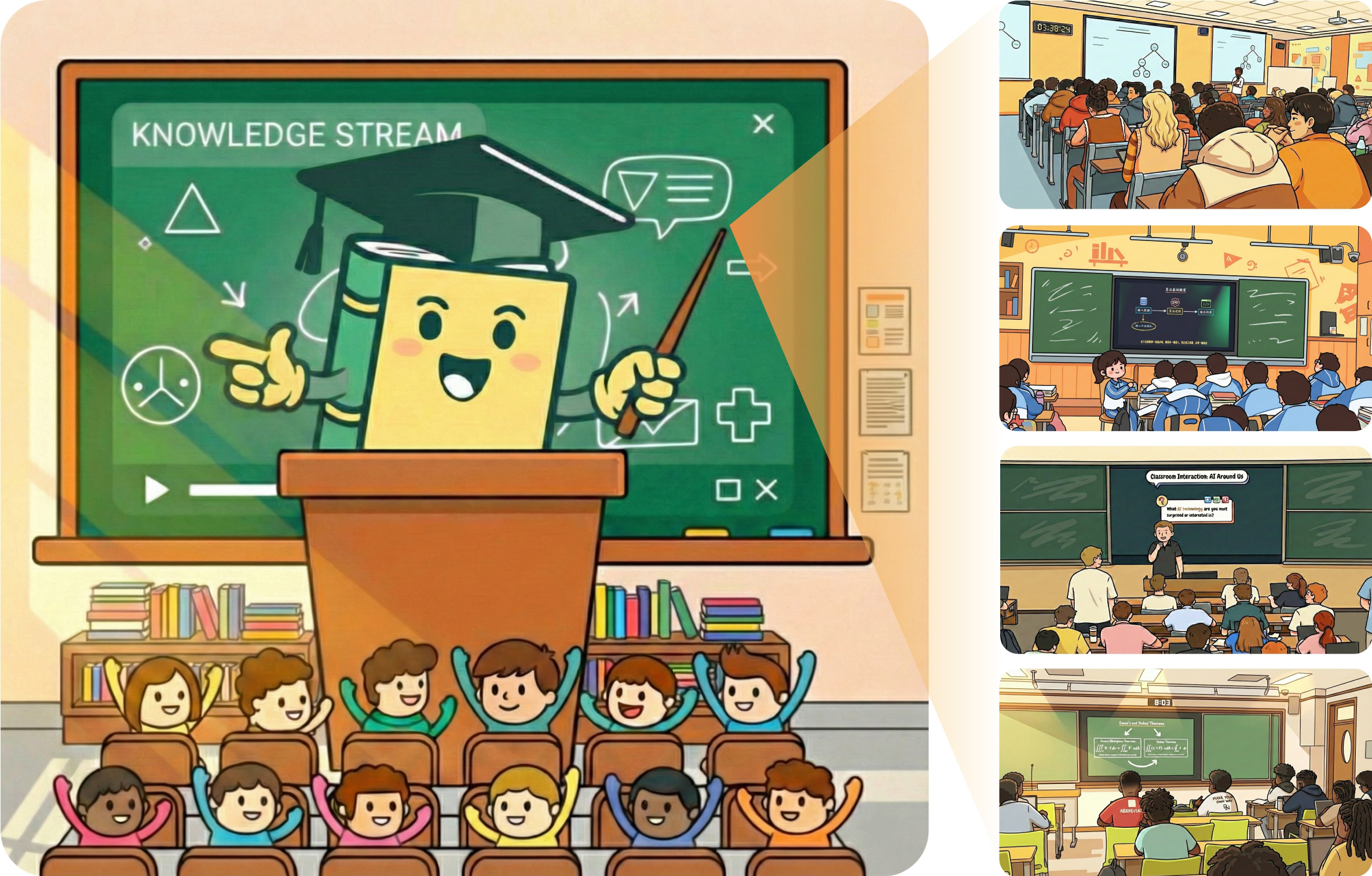}
    \caption{TeachMaster: A code-centric multi-agent framework for Generative Teaching that converts pedagogical intents into ready-to-teach videos.}
    \label{fig:intro}
\end{figure}

To address these limitations, we introduce "\textit{Generative Teaching}"\footnote{We term this intent-driven paradigm "Generative Teaching," or informally "Vibe Teaching," as it allows educators to focus on pedagogical intents while agents handle execution.}, a novel paradigm where the educator transitions from a manual creator to a high-level director for a suite of specialized generative agents. By merely specifying pedagogical objectives, educators can trigger the autonomous creation of curriculum-ready materials, such as videos. This model abstracts away granular implementation details, shifting the workflow from manual content creation to intent-driven instruction, thereby liberating educators from the burdens of extensive preparation.

 From a technical perspective, while end-to-end (E2E) video generation methods~\cite{XingFCDHXWJ25, ho2022video} offer direct output, they often neglect pedagogical structure, yielding results that are uneditable and computationally intensive~\cite{DBLP:conf/cvpr/Wei0CW0GSYG25,liu2024sora,DBLP:journals/corr/abs-2408-11788}. Conversely, approaches mimicking human software usage~\cite{NiuL0FHLKCW24} rely heavily on large-scale multimodal trajectories~\cite{rawles2023androidinthewild,fan2025appcopilot} and substantial training costs~\cite{DBLP:journals/tvcg/XingXLZZHLCCWSW25,XieZCLZCHCSLLXZ24}.
We argue that pixel-level generation is unnecessary for this domain~\cite{chen2025code2video,DBLP:conf/acl/KuCLSYC25}. Instead, we leverage the semantic reasoning, world knowledge, and generative capabilities of pre-trained foundation models~\cite{chang2024survey, WeiTBRZBYBZMCHVLDF22, naveed2025comprehensive}, proposing a novel workflow that employs code as an intermediate representation~\cite{DBLP:journals/corr/abs-2505-21497,DBLP:journals/corr/abs-2510-05096,DBLP:conf/iccv/SurisMV23}. Unlike opaque black-box models, this programmatic approach~\cite{DBLP:conf/eccv/AvetisyanXHYAPZFHOEMNB24,DBLP:journals/corr/abs-2311-16483,DBLP:conf/icml/GaoMZ00YCN23} ensures content interpretability, modularity, and precise control, satisfying the rigorous requirements for scalable educational content production.

More concretely, we present TeachMaster, a multi-agent framework that accepts a lecture outline as input and automates the E2E production of educational videos. 
Acting as a digital production team, TeachMaster orchestrates a collaborative process where agents responsible for content planning, layout design, animation rendering, and speech synthesis work in concert. This synergy ultimately produces coherent, controllable, and scalable educational content.
Experimental results across multiple languages and disciplines reveal that TeachMaster demonstrates superior efficiency without significantly compromising on quality compared to human-made content, thereby offering a more effective solution in the comprehensive trade-off between quality and production cost.

Our contributions are summarized below:
\begin{itemize}[leftmargin=8pt, label=\textbullet, topsep=0pt, itemsep=0pt]
    \item We propose \textit{Generative Teaching}, a novel paradigm that shifts the educator's role from manual creator to high-level director. By prioritizing pedagogical intents over technical implementation, this paradigm empowers generative agents to autonomously handle lesson planning and educational delivery.
    
    \item To realize this vision, we introduce TeachMaster, a multi-agent framework that utilizes code as an intermediate semantic medium. This approach automates the generation of educational videos, facilitating the scalable production of high-quality, interpretable learning resources.

    \item Extensive experiments and real-world deployment across diverse disciplines validate TeachMaster. Performance metrics and user feedback confirm superior structural coherence and cross-modal alignment. TeachMaster reduces production costs to approximately 0.3\% of traditional expenses, balancing efficiency and quality.
\end{itemize}

\section{Method}

 \begin{figure*}[htbp]
     \centering
     \includegraphics[width = \linewidth]{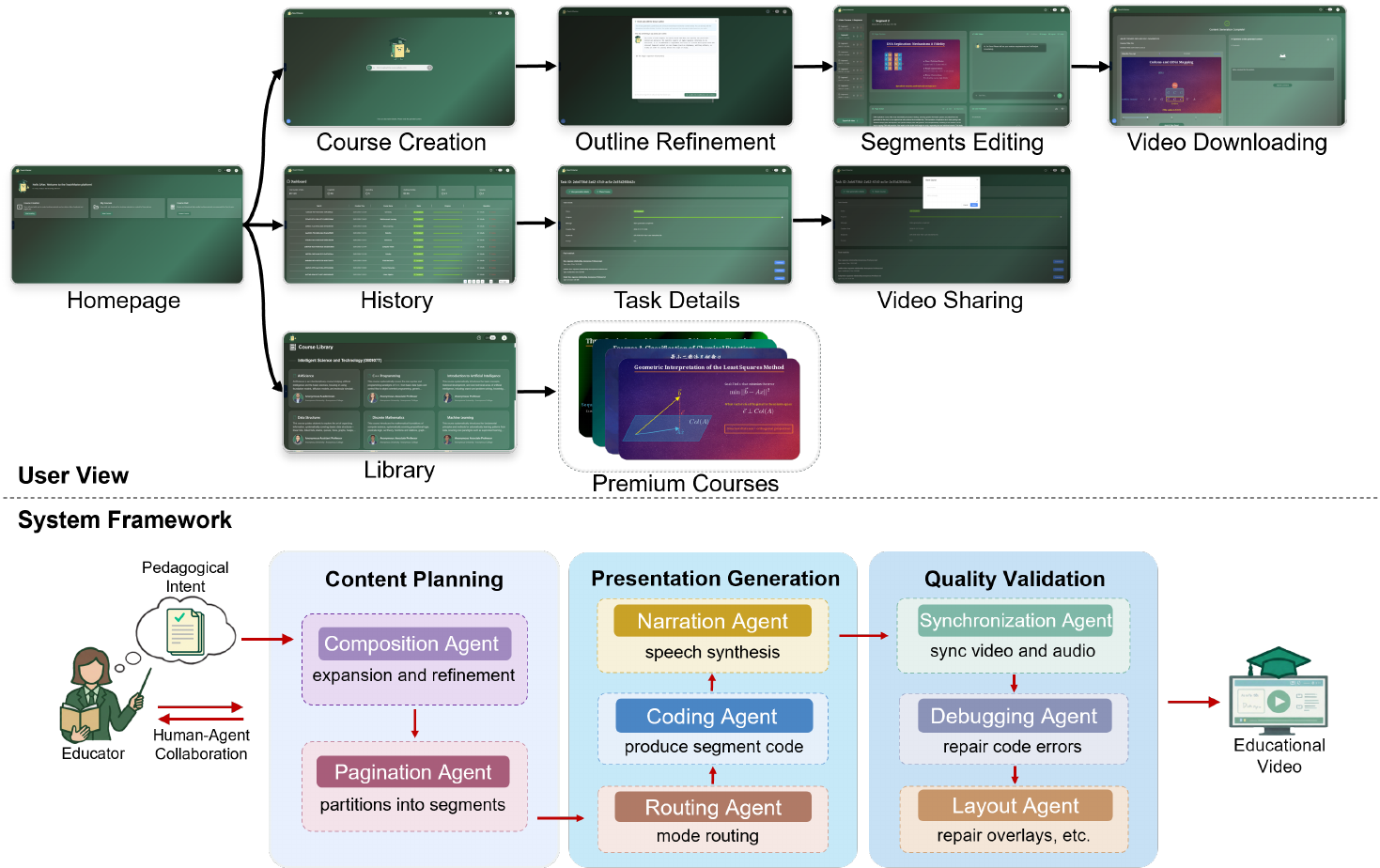}
     \caption{TeachMaster Architecture. The user view (top) offers a seamless platform for course creation and library management, while the underlying framework (bottom) manages a three-stage multi-agent workflow for content planning, presentation generation, and quality validation.}
     \label{fig:framework}
 \end{figure*}

%The core architecture of TeachMaster, as illustrated in Figure~\ref{fig:framework}, consists of a hierarchical multi-agent system using code as a semantic pivot.

%The core architecture of TeachMaster, as illustrated in Figure~\ref{fig:framework}, consists of a hierarchical multi-agent system designed to bridge the gap between abstract lecture outlines and concrete multimodal outputs using code as a semantic pivot.

Generating high-quality educational videos is a multi-stage task that requires navigating a multi-modal transformation from abstract pedagogical intents to intricate content synthesis. Therefore, we introduce TeachMaster, a multi-agent system, as illustrated in Figure~\ref{fig:framework}, that deconstructs the process into three sequential stages: content planning, presentation generation and quality validation. First, the system converts user inputs into page-level blueprints, establishing a semantic foundation. Next, these semantics are transformed into code to produce visual elements, which are then coupled with corresponding narration. Finally, the quality validation ensures the effectiveness of the generated videos through code verification, audio-video synchronization and layout optimization.

For clarity, the generation process can be formalized as follows: given a lecture outline \(k\) (e.g., a set of keywords) and optional configurations \(\Phi\), TeachMaster generates an output set of educational materials. The process is denoted as \( F \):

\vspace{-4pt}
\begin{equation}
\mathcal{O} = F (k, \Phi, f_{\text{human}}) 
\end{equation}

The output set $\mathcal{O} = \{V_{\text{out}}, L_{\text{out}}\}$ contains the generated video and lecture scripts.

\subsection{Content Planning}

%To ensure semantic coherence, the Content Planning Engine transforms raw inputs into structured blueprints. First, the \textit{Content Composer} generates a manuscript $L_{\text{out}}$ by cascading semantic skeletonization ($S$), content expansion ($E$), and length refinement ($R$):

Aiming for high coherence and informativeness, a \textit{composition agent} transforms raw inputs into a comprehensive manuscript $L_{\text{out}}$. Inspired by the communicative reflection in ChatDev~\cite{qian2024chatdev}, it cascades semantic skeletonization ($S$), content expansion ($E$) and length refinement ($R$), where the latter iteratively rectifies the content to match the target video duration $t$:
\vspace{-3pt}

\begin{equation}
L_{\text{out}} = R\Big(E\big(S(k)\big), t\Big)
\end{equation}

Subsequently, a \textit{pagination agent} segments $L_{\text{out}}$ into discrete units $\{P_i\}$. To handle long-text constraints, we employ a Chain-of-Agents (CoA) framework~\cite{Zhang0CPZA24} that partitions $L_{\text{out}}$ into segments $\{D_i\}$ for collaborative processing and aggregation:
\vspace{-5pt}
\begin{equation}
F_{\text{CoA}}(L_{\text{out}}) = \bigcup_{i=1}^{n} F_{i}(D_i) = \{P_i\}_{i=1}^{n}
\end{equation}
where $F_i$ represents a local \textit{pagination agent}.

% To mitigate the semantic fragmentation and coherence issues inherent in direct video generation, TeachMaster implements a dedicated Content Planning Engine. This engine converts raw instructional inputs into context-aware, page-level structured blueprints that preserve conceptual dependencies and reasoning flow.

% The \textit{Content Composer} initially operates on the outline $k$ to generate a comprehensive manuscript $L_{\text{out}}$. This workflow proceeds through three stages: semantic skeletonization ($S$) for concept extraction, content expansion ($E$) for enriching explanations, and global refinement ($R$) for calibrating script length to the target video duration $t$:
% \begin{equation}
%     L_{\text{out}} = f_{\text{c}}(k, t) = R\Big(E\big(S(k)\big), t\Big)
% \end{equation}

% Subsequently, the \textit{Paginator} structures the temporal flow of $L_{\text{out}}$ into discrete units $P_i$, balancing information density and visual complexity. To overcome context limitations in long-text processing, we adopt a Chain-of-Agents (CoA) framework~\cite{Zhang0CPZA24}. This framework partitions the manuscript into segments $D_i$, which are processed by local agents and collaboratively aggregated into a unified sequence:
% \begin{equation}
%     F_{\text{CoA}}(L_{\text{out}}) = \bigcup_{i=1}^{n} F_{i}(D_i)
% \end{equation}
% where $F_{i}$ denotes the operation of a local pagination agent.

\subsection{Presentation Generation}

Presentation generation transforms semantics into multimodal expressions. Unlike unified streams struggling to balance visual reasoning and narrative delivery~\cite{bandyopadhyay-etal-2024-enhancing-presentation}, TeachMaster decouples the process through a dual-stream mechanism. By integrating code-driven visual synthesis with context-aware audio generation, it achieves precise control over visuals and pacing.

%Following content planning, the presentation generation stage transforms structured teaching semantics into interpretable multimodal expressions. Effective teaching requires both well-structured visual reasoning and smooth narrative delivery, which are difficult to achieve through a unified generative stream. Therefore, TeachMaster adopts a dual-stream presentation generation mechanism that combines code-driven visual synthesis with context-aware audio generation to ensure precise control over visuals, reasoning, and pacing.

To bridge abstract semantics and visual representation, a \textit{routing agent} selects the optimal synthesis path for each blueprint $P_i$. While abstract logic is compiled by a standard \textit{coding agent}, content requiring photorealistic assets beyond geometric primitives triggers an \textit{image-enhanced coding agent} to integrate synthesized images as programmable objects. This process is defined as:

%To bridge abstract semantics and visual representation, a \textit{Routing Expert} selects the optimal synthesis path for each blueprint $P_i$. While abstract logic or geometric derivations are compiled directly by a \textit{standard Code Generator}, pages requiring high-fidelity elements trigger an \textit{Image-Enhanced Generator} to integrate synthesized images as programmable objects. This unified process is defined as:

%To bridge the gap between symbolic reasoning and visual perception, TeachMaster employs a hybrid programmatic generation paradigm guided by a \textit{Routing Expert}. For each page blueprint $P_i$, the routing expert first evaluates the semantic requirements to determine the optimal synthesis path. If the content involves abstract logic or geometric derivations, the system invokes a standard \textit{Code Generator} to compile semantics directly into executable visual code. 

%For pages requiring high-fidelity realistic elements, the expert redirects the workflow to an \textit{Image Engine-enhanced generator}. This module first synthesizes realistic visual assets via a generative image model and subsequently integrates these assets as programmable objects into the visual code. The unified generation process is formally defined as:
\vspace{-3pt}
\begin{equation}
    C_i = F_{\text{vis}}(P_i, \text{Mode})
\end{equation}
where $\mathrm{Mode}\in \{\mathrm{Standard},\mathrm{ImageEnhanced}\}$. 
%In practice, the image-enhanced path is selected when a page requires photorealistic assets that are difficult to express with geometric primitives alone. 

%Unlike conventional pixel-based video generation, this code-centric method affords deterministic control over object hierarchies and temporal pacing. Crucially, the final code $C_i$ explicitly encodes the temporal logic of visual events, facilitating downstream synchronization and consistency checks.

Building upon the visual synthesis, a \textit{narration agent} module synthesizes the linguistic component by conditioning on the current page unit $P_i$, the preceding lecture script $T_{i-1}$, and the corresponding visual code $C_i$. This cross-modal reference ensures continuity in terminology and maintains high visual-verbal symmetry, formally defined as:
\begin{equation}
    T_i = F_{\text{narr}}(P_i, T_{i-1}, C_i)
\end{equation}
Subsequently, a \textit{Text-to-Speech (TTS) agent} $F_{\text{tts}}$ processes the script $T_i$ to generate the audio track $A_i$ while simultaneously quantifying speaking rate $r_i$ required for downstream temporal alignment:
\begin{equation}
    A_i, r_i = F_{\text{tts}}(T_i)
\end{equation}

%The derived rate $r_i$ serves as a critical metric for the downstream rhythm optimization module to ensure precise temporal alignment.

\subsection{Quality Validation}
Even with successful multimodal generation, automatically produced educational materials may exhibit execution instability, timing misalignment or visually distracting layouts. To ensure pedagogical quality and structural reliability, TeachMaster incorporates three specialized agents to provide multi-level quality validation: a \textit{debugging agent}, a \textit{synchronization agent}, and a \textit{layout agent}.

To address potential syntax or runtime errors inherent in generative code, the \textit{debugging agent} employs an iterative render-and-repair loop. Upon detecting a rendering failure, the system extracts the error trace and prompts the agent to rectify the specific code segment: 
\begin{equation}
    C_i^{\text{debug}} = F_{\text{debug}}\Big(C_i, \text{Error}(C_i)\Big)
\end{equation}
If the failure persists beyond a retry threshold $\tau$, a fallback mechanism activates, replacing complex elements with standardized templates to improve production stability. 

Subsequently, the \textit{synchronization agent} ensures temporal coherence by aligning visual dynamics with the linguistic flow. Leveraging the event anchors in the debugged code $C_i^{\text{debug}}$ and the semantic units in the script $T_i$, the agent utilizes the calculated speaking rate $r_i$ to determine precise trigger timestamps. It then injects temporal control logic (e.g., waiting statements) into the code:
\begin{equation}
    C_i^{\text{sync}} = F_{\text{sync}}(C_i^{\text{debug}}, T_i, r_i)
\end{equation}
This programmatic adjustment ensures both traceability and reversibility.

The \textit{layout agent} then resolves visual clutter and occlusion by iteratively rectifying geometric conflicts. It first detects overlaps $O_i$, computes optimal coordinates $\Omega_i$ via a heuristic scanning order, and subsequently executes code-level adjustments:
 \begin{equation} 
 \begin{aligned} 
O_i &= F_{\text{detect}}(C_i^{\text{sync}}),\\
 \Omega_i &= F_{\text{retrieve}}({O}_i, \text{dir}_{h,v}), \\ 
 C_i^{\text{layout}} &= F_{\text{layout}}(C_i^{\text{sync}}, \Omega_i)
 \end{aligned} 
 \end{equation}

To bridge automated efficiency with professional preference, the system incorporates a comprehensive human-in-the-loop interface offering two interaction schemes: (1) \textit{Interactive Refinement}, enabling natural language-driven code modifications; and (2) \textit{Direct Code Editing} for granular manual code intervention.
\vspace{-3pt}
\begin{equation}
      C_i^{\text{final}} = F_{\text{human}}(C_i^{\text{layout}}) 
\end{equation}

Afterwards, TeachMaster renders each page into a video segment and merges it with the narration audio to produce the complete video: 
\vspace{-2pt}
\begin{equation}
     V_{\text{out}} = \bigcup_{i=1}^n \text{Render}(C_i^{\text{final}})
\end{equation}

\section{Evaluation}
\label{sec:Evaluation}

\begin{table*}[t]
\centering
\small
\setlength{\tabcolsep}{4pt}
\renewcommand{\arraystretch}{1.1}
\begin{tabular}{l ccccc c @{\hskip 12pt} | ccc}
\toprule
\multirow{2}{*}{\textbf{Method}} & \multicolumn{6}{c}{\textbf{Quality}} & \multicolumn{3}{c}{\textbf{Efficiency}} \\
\cmidrule(lr){2-7} \cmidrule(l){8-10}
& \textbf{Spat.} & \textbf{Rich.} & \textbf{Logic.} & \textbf{T-I Corr.} & \textbf{Acc.} & \textbf{Overall} $\uparrow$ & \textbf{P.T.} $\downarrow$ & \textbf{Dur.} $\uparrow$ & \textbf{Ratio} $\downarrow$ \\
\midrule
\rowcolor{gray!10}
Human              & \textbf{8.22} & \textbf{7.31} & \textbf{8.38} & \textbf{8.29} & \textbf{9.24} & \cellcolor{red4}\textbf{8.29}\phantom{$^\dagger$} & 795.00 & 32.50 & \cellcolor{red1}24.46 \\
Sora 2               & 7.36 & 6.36 & 7.55 & \underline{7.64} & 8.96 & \cellcolor{red1}7.57\phantom{$^\dagger$} & \textbf{3.20} & 0.25 & \cellcolor{red2} 12.80 \\
\rowcolor{gray!10}
TeachMaster$_{\text{Gemini}}$        & \underline{7.97} & \underline{6.98} & \underline{7.97} & 7.63 & \underline{8.99} & \cellcolor{red3}\underline{7.91}$^\dagger$ & \underline{88.43} & \textbf{35.97} & \cellcolor{red4} \textbf{2.46} \\
TeachMaster$_{\text{Qwen}}$        & 7.42 & 6.42 & 7.49 & 7.66 & 8.94 & \cellcolor{red2}7.59$^\dagger$ & 112.80 & \textbf{32.55} & \cellcolor{red3} \underline{3.47} \\
\bottomrule
\end{tabular}
\vspace{2pt}
\caption{
    Video Generation Quality \& Efficiency Evaluation.
    Quality metrics include: 
    Spat. (Spatial Clarity and Layout),
    Rich. (Visual Richness),
    Logic. (Pedagogical and Narrative Logic),
    T-I Corr. (Text–Image Correspondence), and
    Acc. (Factual Accuracy).
    The Efficiency section compares: 
    P.T. (Total Production Time, mins), 
    Dur. (Total Video Duration, mins), and 
    Ratio (Production Time divided by Duration), which indicates the time cost required to produce one minute of content. \textbf{Bold} and \underline{underline} denote the best and the second-best results, respectively.$\dagger$ indicates significant statistical differences ($p \leq 0.05$) between a baseline and ours.
}
\label{tab:video_quality}
\end{table*}

\begin{table*}[t]
    \centering
    \begin{minipage}[t]{0.48\textwidth}
        \centering
        \small
        \setlength{\tabcolsep}{3pt}
        \renewcommand{\arraystretch}{1.1} 
        \begin{tabular}{l cccc c} 
        \toprule
        \multirow{2}{*}{\textbf{Method}} & \textbf{Struct.} & \multicolumn{3}{c}{\textbf{Content Metrics}} & \multirow{2}{*}{\textbf{Overall} $\uparrow$} \\
        \cmidrule(lr){2-2} \cmidrule(lr){3-5}
        & \textbf{Coh.} & \textbf{Acc.} & \textbf{Comp.} & \textbf{Cons.} & \\
        \midrule
        Human    & \textbf{8.90}\phantom{$^\dagger$} & \textbf{9.11}\phantom{$^\dagger$} & \underline{9.05}\phantom{$^\dagger$} & \textbf{8.32}\phantom{$^\dagger$} & \cellcolor{red3} 8.84\phantom{$^\dagger$} \\
        Sora 2       & 3.14\phantom{$^\dagger$} & 6.57\phantom{$^\dagger$} & 1.86\phantom{$^\dagger$} & 6.00\phantom{$^\dagger$} & \cellcolor{red1} 4.39\phantom{$^\dagger$} \\
        \rowcolor{graybg} 
        TeachMaster$_{\text{Gemini}}$ & \underline{8.89}\phantom{$^\dagger$} & \underline{9.00}\phantom{$^\dagger$} & \textbf{9.67}\phantom{$^\dagger$} & \underline{8.22}\phantom{$^\dagger$} & \cellcolor{red4} \textbf{8.95}$^\dagger$ \\
        TeachMaster$_{\text{Qwen}}$ & 8.50\phantom{$^\dagger$} & \underline{9.00}\phantom{$^\dagger$} & 8.17\phantom{$^\dagger$} & 7.67\phantom{$^\dagger$} & \cellcolor{red2} \textbf{8.34}$^\dagger$\\
        \bottomrule
        \end{tabular}
        \vspace{2pt}
        \caption{Educational Script Quality Evaluation. Comparison of script generation performance. Metrics: Coh. (Narrative Coherence), Acc. (Accuracy), Comp. (Completeness), Cons. (Consistency).}
        \label{tab:script_quality}
    \end{minipage}
    \hfill
    \begin{minipage}[t]{0.48\textwidth}
        \centering
        \small
        \setlength{\tabcolsep}{5pt}
        \renewcommand{\arraystretch}{1.2}
        \begin{tabular}{l ccc c}
        \toprule
        \textbf{Method} & \textbf{Cov.} & \textbf{Acc.} & \textbf{Sym.} & \textbf{Overall} $\uparrow$ \\
        \midrule
        \rowcolor{gray!10}
        Human & 8.17\phantom{$^\dagger$} & 7.94\phantom{$^\dagger$} & 8.28\phantom{$^\dagger$} & \cellcolor{red2}8.13\phantom{$^\dagger$} \\
        Sora 2 & 6.64\phantom{$^\dagger$} & 6.59\phantom{$^\dagger$} & 6.73\phantom{$^\dagger$} & \cellcolor{red1}6.65\phantom{$^\dagger$} \\
        \rowcolor{gray!10}
        TeachMaster$_{\text{Gemini}}$ & \underline{8.63}\phantom{$^\dagger$} & \underline{8.11}\phantom{$^\dagger$} & \underline{8.57}\phantom{$^\dagger$} & \cellcolor{red3}\underline{8.44}$^\dagger$ \\
        TeachMaster$_{\text{Qwen}}$ & \textbf{8.93}\phantom{$^\dagger$} & \textbf{8.57}\phantom{$^\dagger$} & \textbf{8.87}\phantom{$^\dagger$} & \cellcolor{red4}\textbf{8.79}$^\dagger$ \\
        \bottomrule
        \end{tabular}
        \vspace{4pt}
        \caption{Cross-modal Semantic Alignment Evaluation. We evaluate performance across three dimensions: Semantic Coverage (Cov.), Referential Accuracy (Acc.), and Visual–Verbal Symmetry (Sym.), which measures the coordination between visual and auditory channels.}
        \label{tab:cross_modal_alignment}
    \end{minipage}
\end{table*}

\noindent \textbf{Metrics.} To evaluate the effectiveness of TeachMaster, we established a comprehensive framework encompassing three primary dimensions: \textit{Video Generation Quality}, \textit{Educational Script Quality}, and \textit{Cross-modal Semantic Alignment}~\cite{huang2024vbench,brame2016effective}. Specifically, we assessed educational videos for visual clarity and pedagogical logic, validated the structural coherence and factual accuracy of the educational scripts, and measured semantic consistency between visual and textual modalities. For quantitative evaluation, we followed established VLM-based assessment paradigms for open-ended generation tasks~\cite{DBLP:conf/emnlp/LiuIXWXZ23,DBLP:conf/naacl/FuNJ024,zheng2023judging}, utilizing GPT-5.2 to score all metrics on a 1–10 scale. We conducted a human-AI preference study involving a panel of three human experts across 300 randomly sampled videos. The overall agreement rate between the experts and GPT-5.2 reached 81.71\%, demonstrating its validity as a scalable proxy for large-scale evaluations.

\noindent \textbf{Baselines.}
We benchmark against two distinct references: (1) E2E: Represented by Sora 2~\cite{openai_sora2_2025}, a state-of-the-art video generation model capable of autonomously producing full educational content. We use it to demonstrate how TeachMaster’s code-centric paradigm ensures pedagogical structure. (2) Human-Crafted: Represented by professional educational videos\footnote{We curated a dataset of highly-rated educational videos and their official captions from publicly available online video-sharing platforms to serve as high-quality human references for both visual and script quality.}, serving as the gold standard.

We exclude existing AI tutoring and slide generation systems~\cite{dan2023educhatlargescalelanguagemodelbased, DBLP:conf/ivsp/DaoLN21} due to a fundamental functional gap. They lack the multimodal synchronization or production scale of TeachMaster. Unlike recent educational agents~\cite{chen2025code2video, DBLP:conf/acl/KuCLSYC25} that focus on short and silent clips, TeachMaster is the first to provide a full-scale, editable solution for industrial-grade course production.

\noindent \textbf{Implementation Details.} 
For visual synthesis, the system renders animations from executable Python scripts via the Manim engine~\cite{manim2025code}. Within the coding agent, TeachMaster features a switchable dual-engine configuration: it either invokes Gemini-3~\cite{team2023gemini} via official APIs or leverages a locally deployed Qwen3-32B~\cite{yang2025qwen3} specifically optimized for high-fidelity Manim code generation. The TTS agent is Minimax~\cite{zhang2025minimax}.

To bolster the performance of Qwen3-32B, we curated a specialized dataset comprising 3735 high-quality human-annotated pairs, categorized by difficulty. Training is underpinned by a curriculum learning strategy~\cite{wang2021survey}. The model was trained on a computing cluster equipped with 8 $\times$ NVIDIA A800 (40GB) GPUs. Fine-tuning was implemented via LoRA ($r=128, \alpha=256$)~\cite{hu2022lora} and DeepSpeed ZeRO-3~\cite{DBLP:conf/kdd/RasleyRRH20} with a learning rate of $1 \times 10^{-5}$.

\subsection{Qualitative Analysis}

As shown in Table~\ref{tab:video_quality}, both Gemini and Qwen powered variants of TeachMaster significantly outperform the E2E baseline and approach the quality of Human references. The generated videos exhibit superior spatial organization and visual balance, ensuring logical consistency that is crucial for education. Notably, TeachMaster supports flexible duration to meet diverse educational needs, whereas E2E models are constrained to short, uncontrollable clips that often fail to deliver comprehensive educational content.

Besides, TeachMaster achieves good performance in script quality, surpassing all baselines (Table~\ref{tab:script_quality}). The generated scripts are logically coherent and pedagogically rigorous, covering essential knowledge points without redundancy. 

A key advantage of our code-centric approach is evident in cross-modal alignment, where TeachMaster outperforms both E2E and Human baselines (Table~\ref{tab:cross_modal_alignment}). It excels in semantic coverage and referential accuracy, ensuring zero information loss between modalities. The high visual–verbal symmetry score demonstrates precise coordination between visual and auditory channels, reinforcing learner perception more effectively than even human-curated content.

In terms of efficiency, TeachMaster demonstrates a decisive advantage. It requires only $\sim$3 minutes to generate one minute of video, which is substantially faster than E2E ($>$12 minutes) and the human baseline. The human time cost is based on established research into educational video production~\cite{hollands2014moocs,g5m2022graphics}. This significant speedup, coupled with high-quality output, highlights TeachMaster’s potential for scalable and automated educational video creation.

\subsection{Practical Deployment and User Feedback}

\begin{figure*}[t]
\centering
\subfigure[Educator perception of preparation time reduction.]{
    \includegraphics[width=0.23\textwidth]{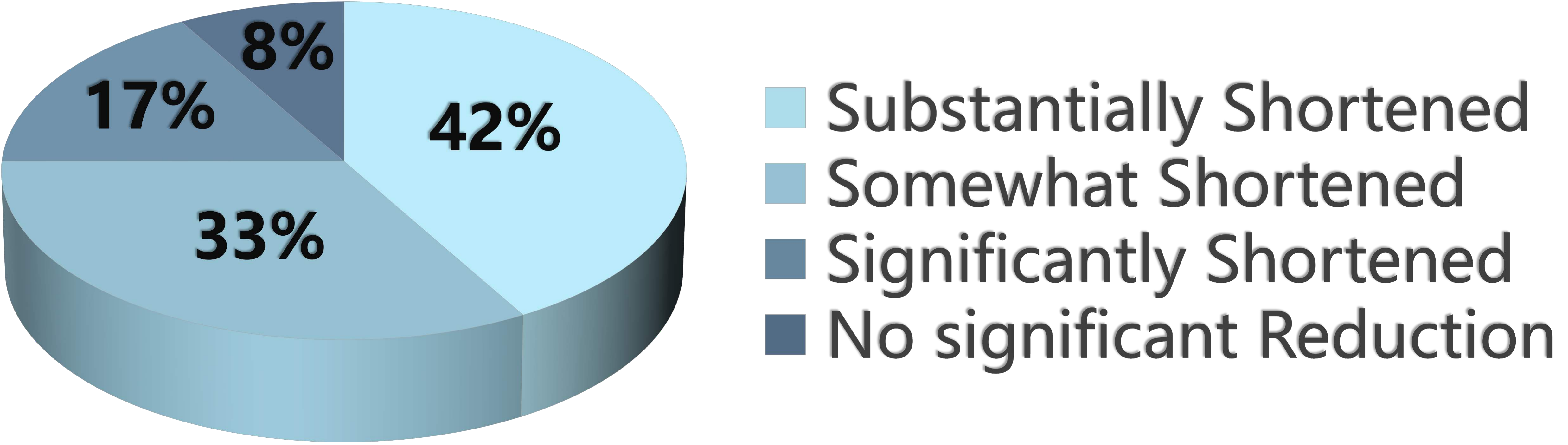}
}
\hfill
\subfigure[Video coverage of the original instructional intent.]{
    \includegraphics[width=0.23\textwidth]{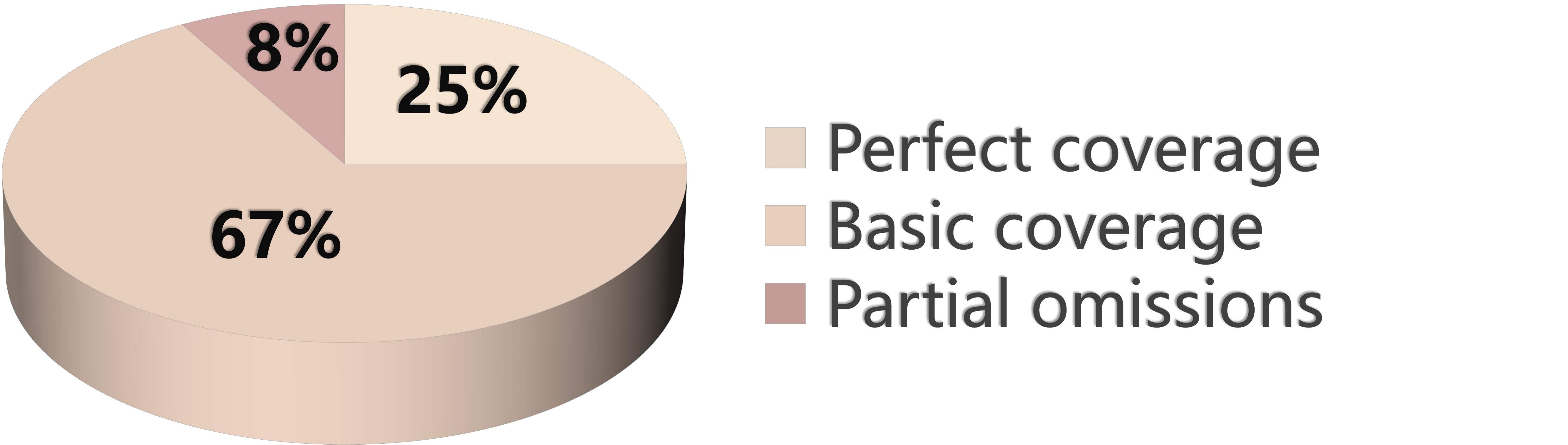}
}
\hfill
\subfigure[Distribution of key advantages of TeachMaster over traditional methods. ]{
    \includegraphics[width=0.23\textwidth]{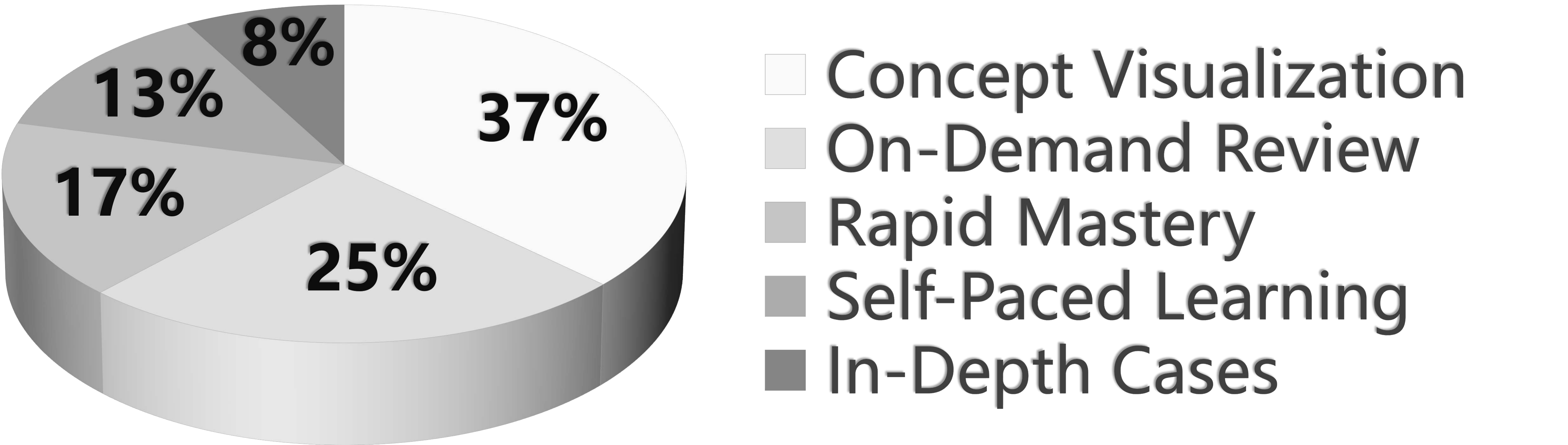}
}
\hfill
\subfigure[Student suggestions for future content optimization.]{
    \includegraphics[width=0.23\textwidth]{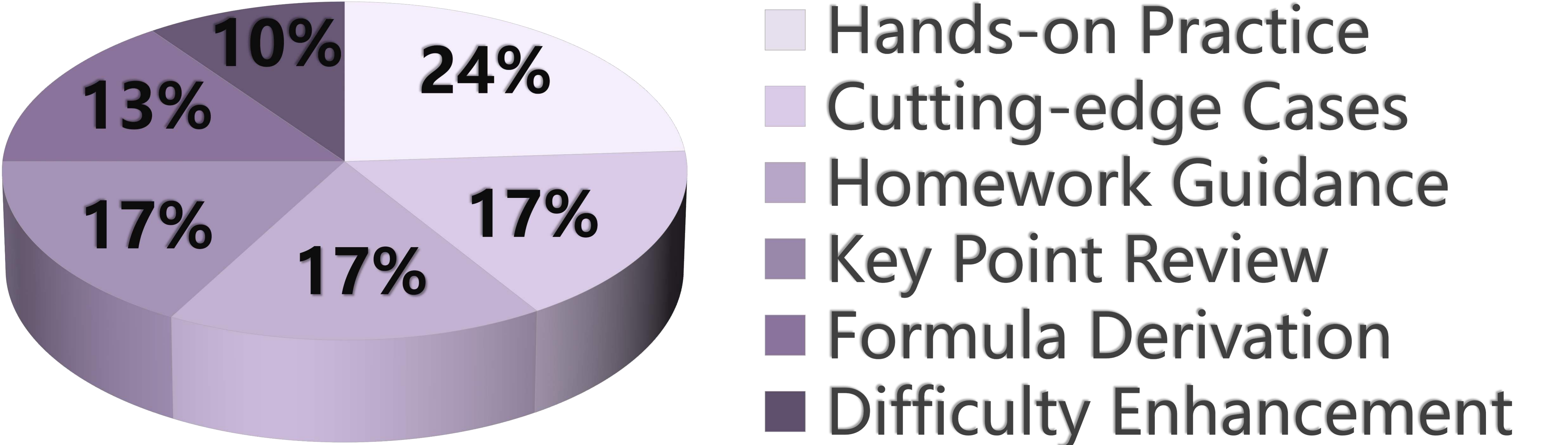}
}
\caption{Analysis of Real-World Classroom Feedback.}
\label{fig:feedback}
\end{figure*}

\begin{figure}[h]
    \centering
    \includegraphics[width=1\linewidth]{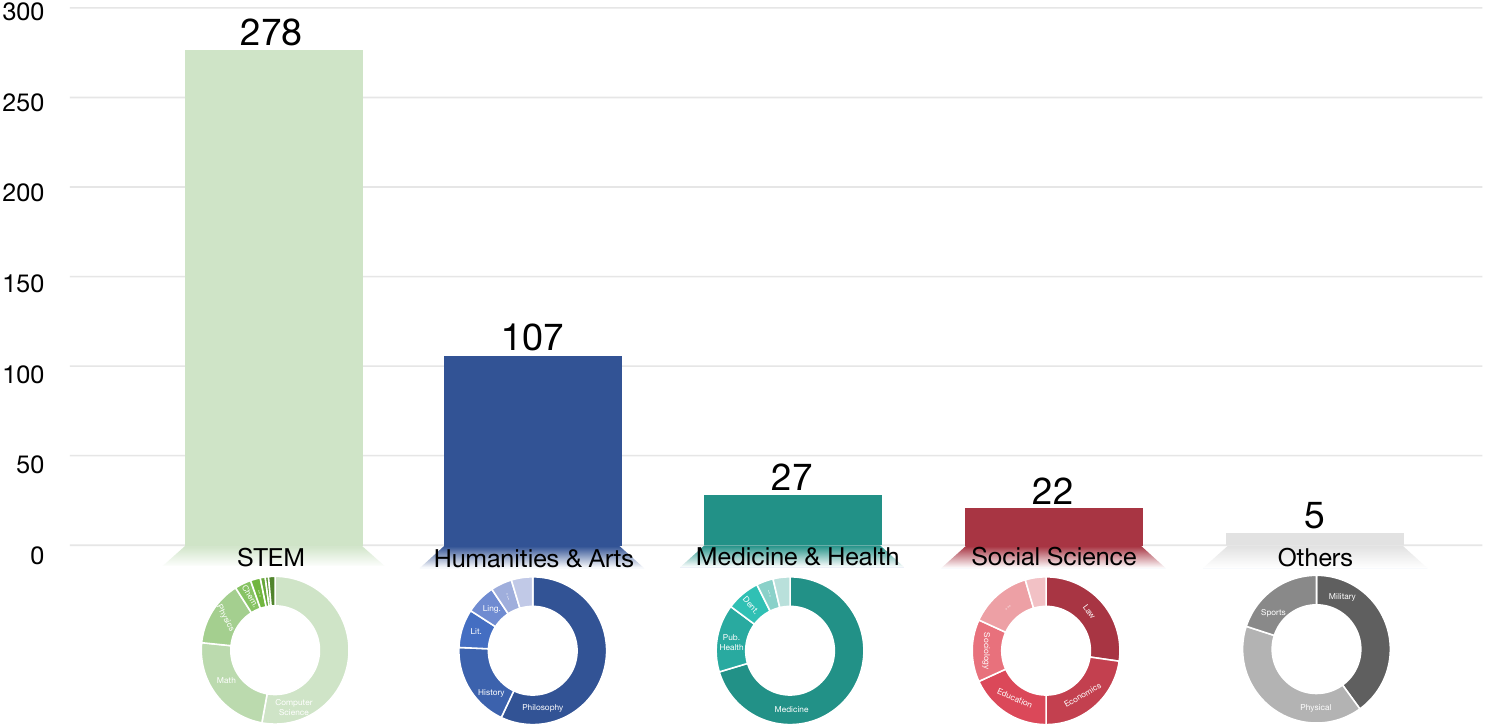}
    \caption{Quantitative distribution of academic materials across major disciplinary categories.}
    \label{fig:distribution}
\end{figure}

\paragraph{Deployment}TeachMaster is in active deployment on a computing cluster equipped with $8 \times $ NVIDIA A800 GPUs (each with 40GB memory). This infrastructure supports seamless switching between dual engines based on user selection. To ensure system stability under multi-user scenarios, we implemented an \textit{asynchronous task queue} mechanism, effectively managing computational resources for simultaneous generation requests. Leveraging this robust foundation, the system has demonstrated exceptional practical performance. As shown in Figure~\ref{fig:intro}, it has served over 1000 educators across primary, vocational, and QS Global Top 50 universities. To date, the platform has generated more than 30,000 minutes of educational content.  Based on cumulative production records spanning more than 40 distinct disciplines (See Figure~\ref{fig:distribution}), TeachMaster demonstrates remarkable cross-disciplinary generalization.

Notably, with this deployment setup, the cost of producing a standard 45-hour semester-long course is approximately \$83.70, a mere 0.3\% of traditional online course production expenses.

\paragraph{User Feedback}As illustrated in Figure~\ref{fig:feedback}, we conducted surveys among both educators and students to evaluate the efficacy of TeachMaster in real-world pedagogical settings.

By prioritizing code-level editability and an automated workflow, TeachMaster significantly reduces preparation time, allowing educators to focus on high-value instructional guidance. Empirical data demonstrates that over 75.2\% of pages require no manual intervention, while the remainder are finalized within an average of just 1.88 interaction rounds. This human-agent collaborative paradigm effectively overcomes the black-box limitations of conventional AI, ensuring pedagogical rigor while maintaining precise control over specialized content. Simultaneously, the system surpasses traditional lectures in both conceptual clarity and visual engagement. Students reported higher comprehension efficiency for abstract knowledge and appreciated the flexibility of self-paced review, while also expressing a growing demand for advanced modules like cutting-edge case studies and exam-tailored exercises.
%除了描述学生的反馈，还需要老师/用户的使用反馈

% \begin{figure*}[htbp]
% \centering

% \subfigure[Abstract Algebra]{
%     \includegraphics[width=0.45\textwidth]{figs/demo1cn.pdf}
% }
% \hfill
% \subfigure[Quantum Physics]{
%     \includegraphics[width=0.45\textwidth]{figs/demo2cn.pdf}
% }

% \vspace{4pt}

% \subfigure[Supervised Learning]{
%     \includegraphics[width=0.45\textwidth]{figs/demo3cn.pdf}
% }
% \hfill
% \subfigure[Introduction to AI]{
%     \includegraphics[width=0.45\textwidth]{figs/demo4cn.pdf}
% }

% \vspace{4pt}

% \subfigure[Molecular Biology]{
%     \includegraphics[width=0.45\textwidth]{figs/demo1en.pdf}
% }
% \hfill
% \subfigure[Linguistics]{
%     \includegraphics[width=0.45\textwidth]{figs/demo2en.pdf}
% }

% \vspace{4pt}

% \subfigure[Chemistry]{
%     \includegraphics[width=0.45\textwidth]{figs/demo3en.pdf}
% }
% \hfill
% \subfigure[Embodied Intelligence]{
%     \includegraphics[width=0.45\textwidth]{figs/demo4en.pdf}
% }

% \vspace{2pt}
% \caption{Examples of TeachMaster-generated bilingual course materials across multiple disciplines and languages. The system transforms textual outlines into multimodal teaching materials (including animated visuals, narration, voiceovers, and other customizable configurations).}
% \label{fig:teachmaster_examples}
% \end{figure*}
\subsection{Case Study}

\begin{figure*}[t]
\centering

\subfigure[Supervised Learning]{
    \includegraphics[width=0.45\textwidth]{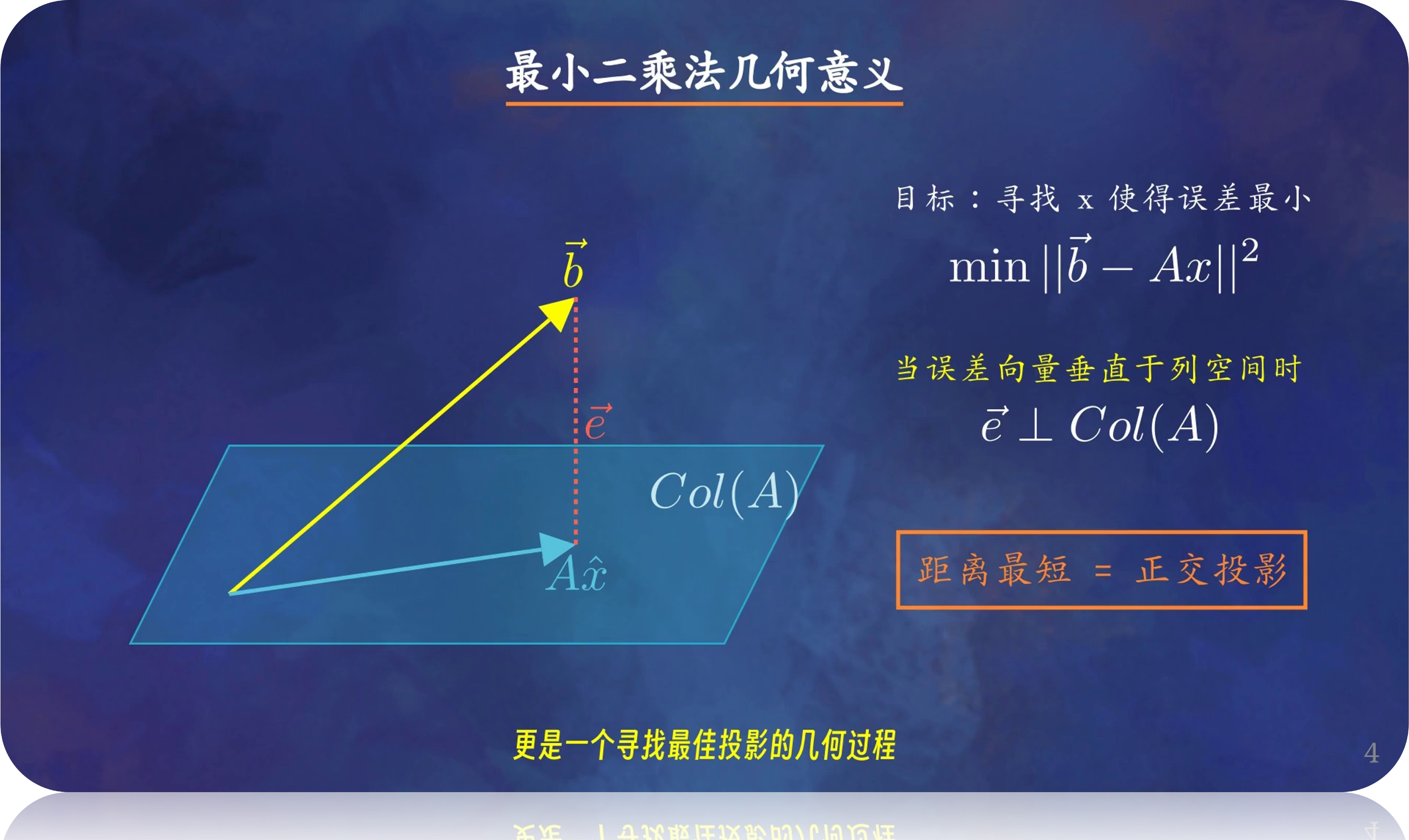}
}
\hfill
\subfigure[Introduction to AI]{
    \includegraphics[width=0.45\textwidth]{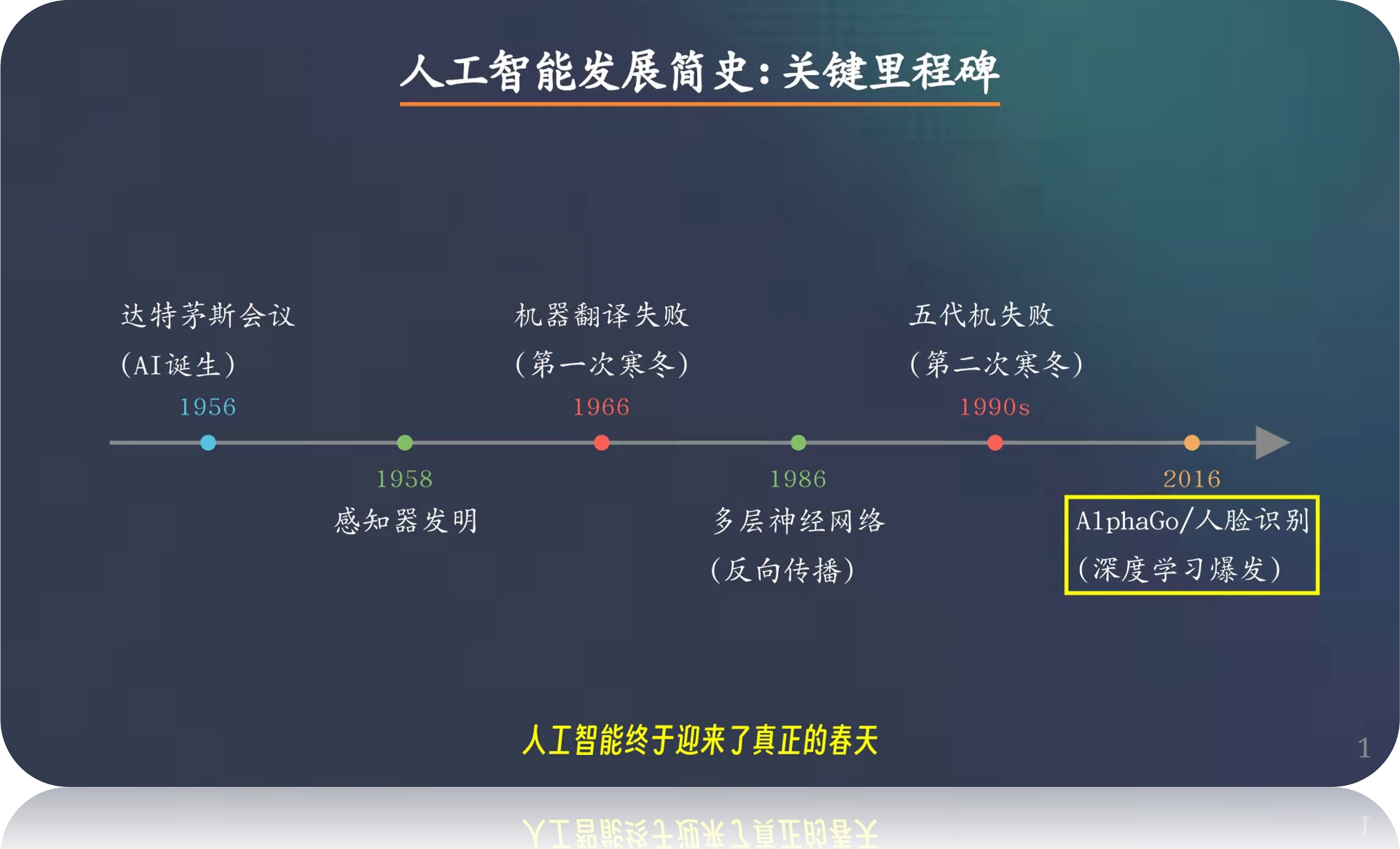}
}

\vspace{4pt}

\subfigure[Molecular Biology]{
    \includegraphics[width=0.45\textwidth]{figs/demo1en.pdf}
}
\hfill
\subfigure[Linguistics]{
    \includegraphics[width=0.45\textwidth]{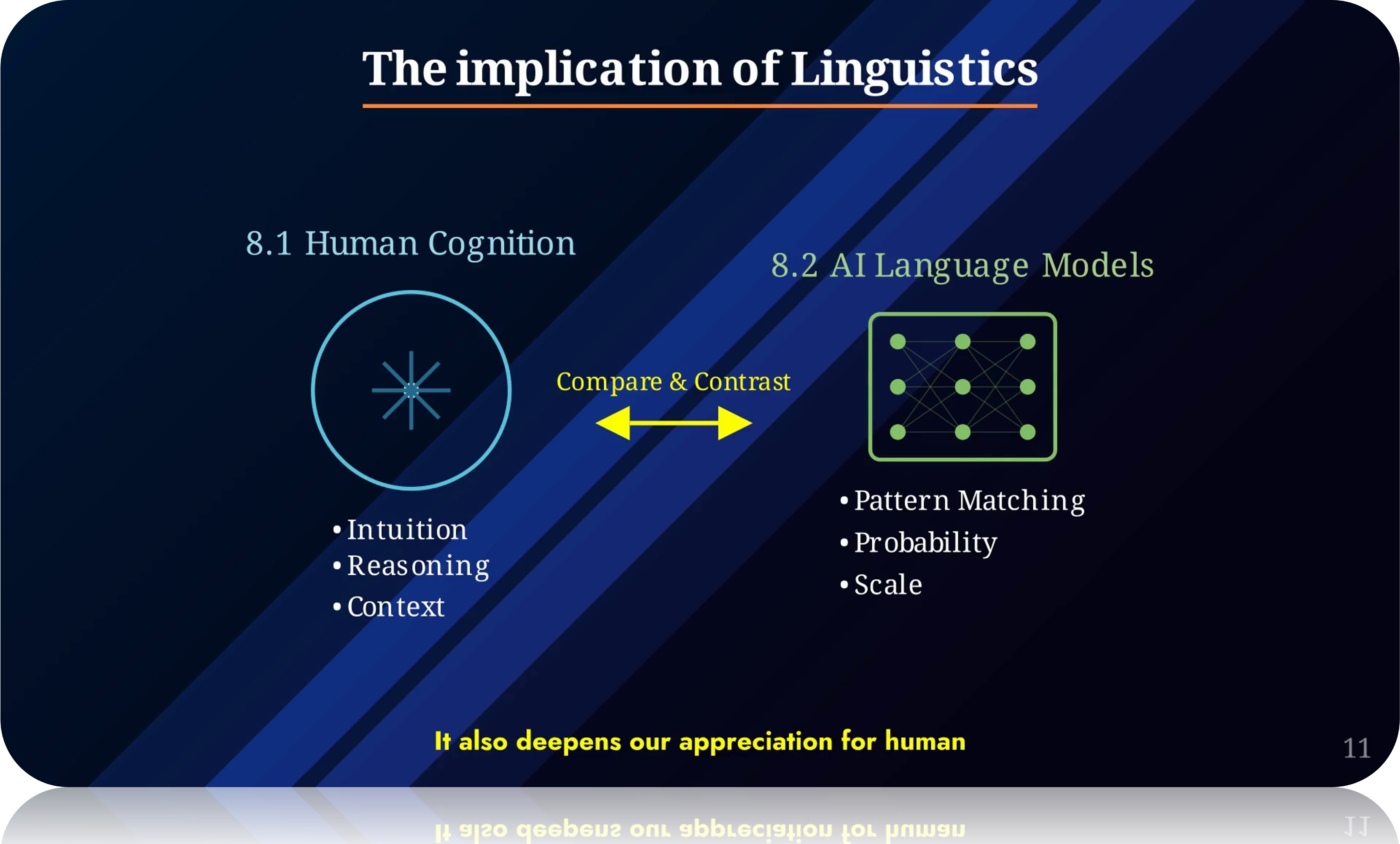}
}

\vspace{2pt}
\caption{Examples of TeachMaster-generated bilingual course materials across multiple disciplines. The system transforms structured textual outlines into complete multimodal teaching materials, including animated visuals, narration, voiceovers.}
\label{fig:teachmaster_examples}
\end{figure*}

Figure~\ref{fig:teachmaster_examples} showcases TeachMaster's versatility through generated video clips across diverse academic disciplines. These cases, combined with real-world applications, confirm the system as a universal teaching assistant that seamlessly adapts to subject-specific requirements.

The system demonstrates robust performance in both Chinese and English contexts, exhibiting precise control over concise text generation, adaptive color coding, high-fidelity imagery, and dynamic animations. For instance, in Supervised Learning, geometric projections demystify the mathematical intuition behind the Least Squares method. Similarly, the system visualizes the transition of protein chains into functional 3D structures, facilitating an intuitive understanding of complex natural sciences.

\section{Related Work}

Large Language Models (LLMs) have been extended from static predictors to autonomous agents capable of planning, tool use and memory~\cite{li-etal-2025-knowledge-boundary}. Early studies primarily focused on single-agent systems for specific tasks~\cite{xi2023rise,zhou2023webarena,huang2022language,DBLP:conf/acl/YangZWCHYLTLYLS24,park2023generative,shinn2023reflexion,bran2023chemcrow,gou2023tora,DBLP:conf/acl/0036PWZSL0YRZ025}. To address the growing complexity of real-world workflows, subsequent work has explored multi-agent collaboration~\cite{wu2024autogen,hong2023metagpt,du2023improving,chen2023agentverse,DBLP:journals/corr/abs-2505-19591,DBLP:conf/acl/DuQLXWQDCYTXH25,NEURIPS2023_a3621ee9}.

Meanwhile, AI-Generated Content (AIGC) has evolved from unimodal generation toward integrated multimodal systems, covering text~\cite{zhao2023survey,DBLP:journals/corr/abs-2105-10311}, code~\cite{DBLP:conf/icml/WangCY0L0J24,DBLP:conf/nips/YangJWLYNP24}, and audiovisual outputs~\cite{DBLP:conf/bigdataconf/WuGCWY23}. Recent systems incorporate verification or refinement mechanisms for text and code generation~\cite{wang2024openhands,gao2025trae}, while diffusion and transformer based models enable high-quality visual synthesis and voice cloning~\cite{ho2020denoising,du2024cosyvoice}, supporting E2E multimedia generation by autonomous agents~\cite{zhang2026versatile}.

In educational contexts, earlier AI systems focused on text-based tasks~\cite{mageira2022educational,grassini2023shaping,dan2023educhatlargescalelanguagemodelbased,DBLP:conf/ivsp/DaoLN21,DBLP:conf/iccv/LeeALNM23,yu2024moocmaicreshapingonline,zhang2024simulatingclassroomeducationllmempowered}. While later works explored visual generation, they relied on rigid templates~\cite{imran2024google}. Recently, agentic frameworks have emerged to automate the production pipeline~\cite{zhangli2024awakingslidestuningfreeknowledgeregulated}.

\section{Conclusion}

We introduce TeachMaster, the pioneering framework for the Generative Teaching paradigm. Its code-centric workflow for synthesizing scripts and dynamic animations ensures transparency and editability while lowering production barriers. Within this framework, we establish a novel human-AI collaboration model: positioning AI as the technical foundation for large-scale multimodal synthesis, while empowering educators as posteriors to exercise high-level logical supervision. This approach achieves a deep complementarity between machine efficiency and human educational intuition.

Large-scale real-world deployment validates the system's exceptional balance between production efficiency and instructional quality, achieving a substantial reduction in production costs compared to traditional online courses. Ultimately, Generative Teaching will serve as a catalyst, liberating educators from repetitive content production to focus on the art of personalized instruction, while AI ensures the scalable expansion and precise presentation of knowledge.

\section*{Acknowledgements}

This work was supported by the Shanghai Municipal Special Program for Basic Research on General AI Foundation Models (Grant No. 2025SHZDZX026D04).

\bibliography{custom}

\end{document}